# A new opportunity for two-dimensional van der Waals heterostructures: making steep-slope transistors

Juan Lu[+a], Jing Pei[+a], Yuzheng Guo[+b], Jian Gong[c], Huanglong Li[a*]

[a]*Department of Precision Instrument, Center for Brain Inspired Computing Research, Tsinghua University, Beijing 100084, P. R. China*

[b]*School of Electrical Engineering and Automation, Wuhan University, Wuhan 430072, P. R. China*

[c]*School of Physics and Technology, Inner Mongolia University, Hohhot 010021, P. R. China*

[+]*Equal contribution*

*E-mail: li_huanglong@mail.tsinghua.edu.cn

**Abstract**

The use of a foreign metallic cold source (CS) has recently been proposed as a promising approach toward the steep-slope field-effect-transistor (FET). In addition to the selection of source material with desired density of states-energy relation (D(E)), engineering the source: channel interface for gate-tunable channel-barrier is crucial to a CS-FET. However, conventional metal: semiconductor (MS)-interfaces generally suffer from strong Fermi-level-pinning due to the inevitable chemical disorder and defect-induced gap states, precluding the gate-tunability of the barriers. By comprehensive materials and device modeling at the atomic-scale, we report that the two-dimensional (2D)-van der Waals (vdW)-MS-interfaces, with their atomic sharpness and cleanness, can be considered as general ingredients for CS-FETs. As test cases, InSe-based n-type FETs are studied. It is found that graphene can be spontaneously p-type doped along with slightly opened bandgap around the Dirac-point by interfacing with InSe, resulting in super-exponentially decaying hot carrier density with increasing n-type channel-barrier. Moreover, the D(E) relations suggest that 2D-transition-metal-dichalcogenides and 2D-transition-metal-carbides are rich libraries of CS materials. Both graphene and H-TaTe2 CSs lead to subthreshold swing below 60 mV/decade. This work broadens the application potentials of 2D-vdW-MS-heterostructures and serves as a springboard for more studies on low-power electronics based on 2D materials.

**Introduction**

Propelling the current interest in electronics based on two-dimensional (2D)

materials is the continual discovery of new physical phenomena and the resultant device innovations.[1-4] Field-effect-transistor (FET) technology has been the workhorse of modern semiconductor industry. 2D semiconducting materials such as $MoS_2$,[5-7] black phosphorus,[8, 9] InSe[10-13] and $PdSe_2$[14] have been widely studied as channel materials to replace bulk silicon because of the ideal electrostatic control of the atomically thin channels. Another unique property of 2D materials is the ability to form vertical heterostructures by van der Waals (vdW) interactions without direct chemical bonding.[15-18] This offers considerable freedom in integrating various materials without the constraints of crystal lattice matching for unprecedented functions. In regard to FET devices, it has been predicted by theory[19] and demonstrated by experiment[20, 21] that vdW metal: semiconductor (MS) source/drain-contacts outperform conventional contacts[22] in achieving low contact resistance. This has been attributed to the weak Fermi-level (FL) pinning effect because of the less chemical disorder and defect-induced gap states at the 2D vdW MS interfaces.

The reinvention of the semiconductor device technology has been catalyzed by the limits of power dissipation. Due to the thermionic limit, conventional FETs require at least 60 mV of gate voltage to increase the source-drain current by one order of magnitude at room temperature, precluding the continuous scaling of the supply voltage and the decrease of power consumption.[23, 24] Various device innovations have been proposed to lower the subthreshold swing (SS) below 60 mV/decade.[25, 26] Tunneling FET (T-FET) by using heterojunction-channel[27, 28] and negative-capacitance FET (NC-FET) by using ferroelectric-gate-oxide[29, 30] are most extensively studied. However, T-FETs have now been well known for low on-current; meanwhile, NC-FETs can often suffer from large hysteresis. Recently, cold source (CS) has emerged as a promising solution to overcome the limitations of T-FETs and NC-FETs. In a CS-FET,[31] the hot carrier (HC) density of states (DOS) of the CS should decrease with increasing channel barrier, which results in a super-exponentially decreasing HC density (n(E)). Therefore, more localized carrier distribution around the FL without a long thermal tail above the channel barrier can be achieved for lower SS.

In addition to the desired DOS-Energy relation (D(E)) of the source material,

source: channel interface with weak FL pinning is crucial to a CS-FET. In fact, the gate-tunability of the channel barrier distinguishes a CS-FET from conventional Schottky-barrier (SB)-FET where the FL of the MS interface is pinned and the barrier height can hardly be tuned by the gate-voltage.

Considering that 2D vdW MS heterostructures show weak FL pinning and enable effective tuning of the SBs, we investigate in this work the feasibility of using such unique heterostructures as general ingredients for CS-FETs. As test cases, monolayer InSe-based n-type high-mobility FETs with prospective 2D CSs are studied. By analyzing the D(E) relation, we find that, in addition to graphene, 2D transition-metal dichalcogenides (TMD) and 2D transition-metal carbides (MXene) are rich libraries of CS materials. Both graphene and H-TaTe$_2$ CSs are found to result in SS below 60 mV/decade, which can be explained by their desired HC distributions and the gate-tunable source: channel barrier heights. This work opens up new opportunities at the confluence of 2D materials and electronics.

**Results and Discussion**

Few-layer InSe-based transistor has recently been experimentally fabricated, exhibiting high electron mobility and air-stability.[10, 32] The high quality of InSe makes it competitive with atomically thin dichalcogenides and black phosphorus. As test cases, monolayer InSe-based transistors are investigated in this work. With the linearly decreasing hot hole DOS, n-type doped graphene has been experimentally demonstrated as an effective CS material for p-type carbon nanotube transistors.[31] Here, we first model a graphene CS-FET, as illustrated in figure 1(a). The graphene source with the length of about 1.2 nm is extended to the gate region. The channel maintains intrinsic and drain region is n-type doped with the concentration of $10^{20}$ cm$^{-3}$. The calculated lattice constants of the hexagonal unit cell of monolayer InSe is a=b=4.09 Å, in agreement with previous results.[33] Because the bandgap calculated by generalized gradient approximation (GGA) functional is underestimated, we use meta-GGA[34] to obtain a value of 1.95 eV which is comparable to the experimental value.[10] Figure 1(b) shows the band structure of InSe: graphene heterostructure calculated by standard periodic supercell method. It can be seen that InSe and graphene form a low n-type SB with the height of 0.04 eV. In addition, it is observed that graphene is spontaneously p-type doped by interfacing with InSe, leading naturally to the desired D(E) relation, namely, decreasing hot electron DOS, for n-type transistor. Moreover, bandgap opening of graphene by about 10 meV occurs in the formation of vdW heterostructure.

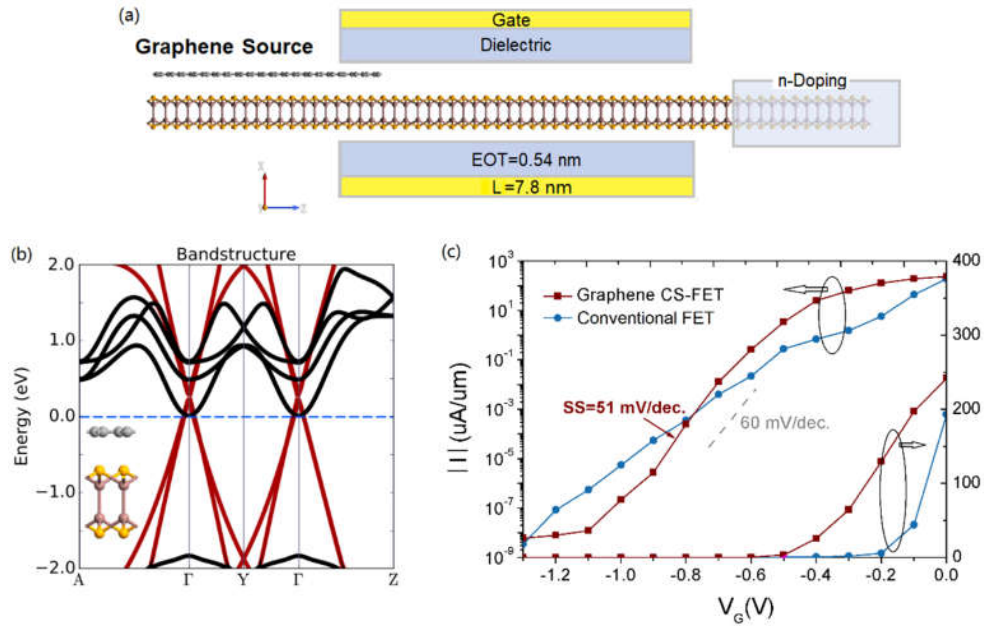

**Fig.1** (a) Side view of n-type InSe-based graphene CS-FET. The equivalent-oxide-thickness (EOT=0.54 nm), gate length (L=7.8 nm) and supply voltage ($V_{SD}$=0.74 V) are parametrized according to the International Technology Roadmap of Semiconductors (ITRS)[35] for high performance transistors in 2021. (b) Band structure of InSe: graphene heterojunction. The electronic states contributed by graphene and InSe are marked by the red and black curves, respectively. The Fermi energy is set to zero. The inset shows the relaxed atomic structure, the yellow, pink and grey balls represent In, Se, and C atoms, respectively. (c) Transfer characteristics of InSe-based FETs with and without graphene source, I-V cures are displayed in both logarithmic and linear scales.

The calculated transfer curve of graphene CS-FET is plotted in figure 1(c). Compared with conventional FET with n-type doped InSe source (with other geometric and electronic parameters the same), it can be seen that graphene CS-FET exhibits much suppressed SS. In particular, the minimum SS value of 51 mV/decade occurs in the gate-voltage ($V_G$) region between -0.9 V and -0.8 V. Moreover, the CS-FET device shows large ON/OFF ratio, exceeding $10^9$ in the $V_G$ sweeping region from -1.1 V to -0.4 V at $V_{SD}$=0.74 V.

To understand the operational principles of the graphene CS-FET, figure 2(a, b) show the device DOSs projected to three regions, i.e., source, channel and drain, in the ON and OFF states of the device, respectively. It is seen that the height of the n-type source-channel SB is gate-tunable with low barrier height for easy electron injection in the ON state and high barrier height to block electron injection in the OFF state. Figure 2(c) shows the transmission spectrums of the device in the ON and OFF states. As expected, there is significant contrast of the transmission probability at the FL of the source region ($E_F(S)$) between the ON and OFF states due to the change of the channel barrier. Figure 2(d) shows the atomic DOSs projected to graphene source, InSe in the channel and InSe in the drain for the ON- and OFF-state devices. In agreement with the previous supercell calculation (figure 1(b)), in the device modelling, the graphene is found to be spontaneously p-type doped along with slightly opened energy gap around the Dirac point. In the process of turning-off, the height of the source-channel SB gradually increases with increasing $V_G$ (in absolute values). Because the FL of graphene source lies in the lower Dirac cone, the hot electron DOS decreases with increasing channel barrier, which results in a super-exponentially decreasing n(E) relation and thus suppressed thermal tail above the potential barrier to the channel. This is supported by the zoom-in subplots of D(E) and n(E) near the $V_G$ where the smallest SS occurs (figure 2(e)). Moreover, a slightly opened bandgap of graphene is observed in figure 2(d), in consistence with the previous supercell calculation (figure 1(b)). This makes graphene CS more effective in suppressing the thermal tail contribution to the OFF-state current for smaller SS,[36] which is evidenced by the concurrence of smallest SS and the coincidence of the conduction band minimum (CBM) of InSe with the energy gap of graphene source.

Hereto, the use of graphene CS to achieve steep-slope InSe-based transistor has been discussed. This has been attributed to the gate-tunable channel barrier and the desired D(E) relation that are provided by the 2D vdW heterostructure.

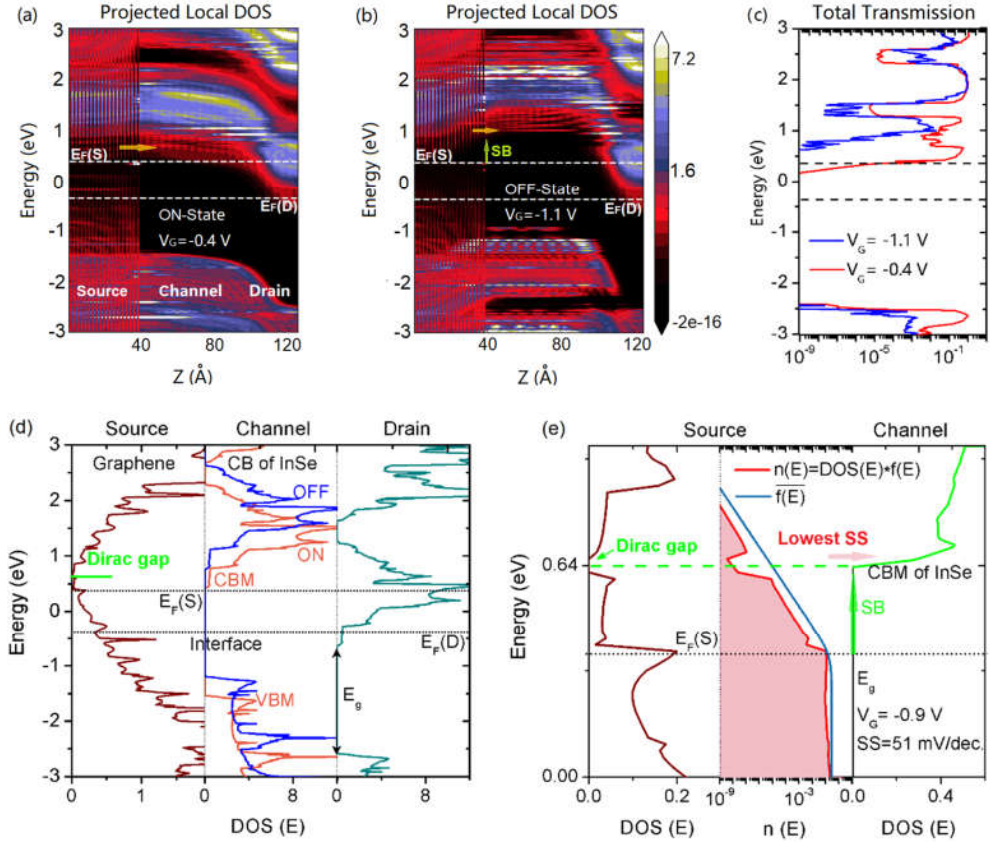

**Fig.2** DOSs projected to graphene CS-FET along the transport direction for (a) ON-and (b) OFF-state devices. (c) Transmission spectrums for ON- and OFF-states. (d) Atomic DOSs projected to graphene source, InSe in the channel and InSe in the drain for the ON- and OFF-state devices. (e) zoomed-in DOS(E) and n(E) in the energy region near which the steepest slope of SS=51 mV/decade occurs. For comparison, the linearly-scaled Boltzmann distribution function $\overline{f(E)}$ that equals n(E) at the FL is shown in blue curve.

Since the discovery of graphene, a wide range of graphene-like 2D vdW materials have been explored and exploited. Here, a natural question arises: do we have other 2D materials options than graphene as the vdW CS in achieving steep-slope transistors? To answer this question, we first study the D(E) and n(E) relations of some selected 2D materials. We consider two classes of 2D materials beyond graphene, i.e., TMDs and MXenes. Recently, it has been reported that certain MXenes are desired electrode materials to achieve low contact resistance for 2D InSe nanodevices.[37] The DOSs and carrier distributions (n(E)) of three monolayer $M_3C_2$-type MXenes, namely, $Zn_3C_2$, $Hg_3C_2$ and $Cd_3C_2$, are shown in figure 3(a-c), respectively. These selected MXenes show p-type semi-metallic characters with Dirac band dispersion, similar to p-type doped graphene. Therefore, the DOS above the FL is a decreasing function of energy, leading to the formation of a super-exponentially decreasing n(E). As aforementioned, this is key to the localization of carrier distribution around the FL and the suppression of thermal tail contribution to the OFF current. For TMDs, the DOSs and n(E) of H-phase metallic $TaTe_2$, $VTe_2$ and $VSe_2$ are shown in figure 3(d-e), respectively. These TMDs have overall decaying DOSs with energies. In addition to these desired D(E) relations, the band structures of H-$VTe_2$ and H-$VSe_2$ have energy gaps at slightly higher energy values, resulting in an abrupt cutoff of the carrier distributions. These properties, along with the nonchemical nature of vdW interaction, could offer certain MXenes and TMDs the opportunities to be CS materials for steep-slope 2D FETs.

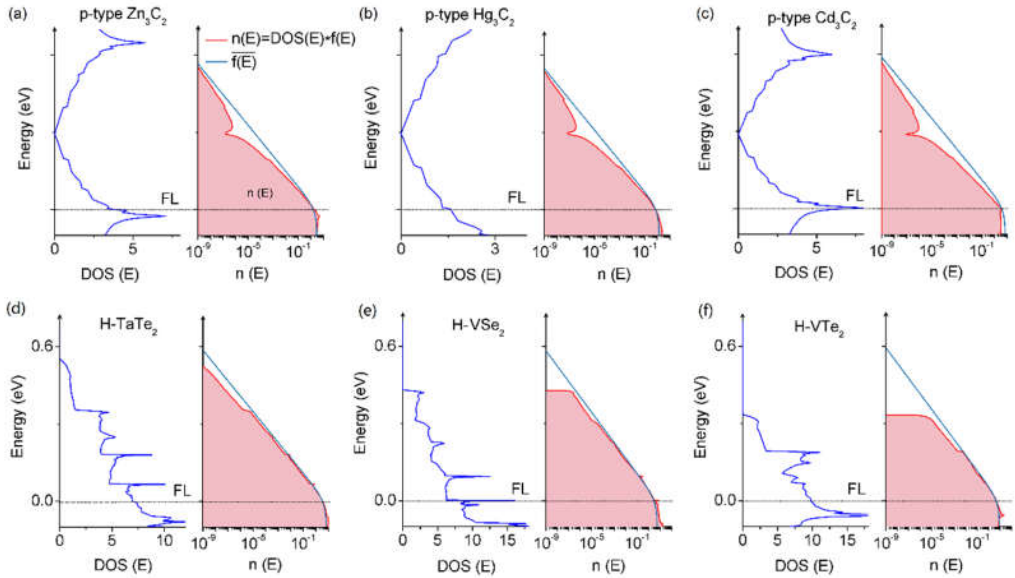

**Fig.3** DOS(E) and n(E) relations of (a) $Zn_3C_2$, (b) $Hg_3C_2$, (c) $Cd_3C_2$, (d) H-$TaTe_2$, (e) H-$VSe_2$, and (f) H-$VTe_2$.

As another test case, we model an H-$TaTe_2$ CS-FET, as shown in figure 4(a). The H-$TaTe_2$ source is extended to the gate region. The geometric and electrical parameters are the same as those of the graphene CS-FET. Figure 4(b) shows the band structure of InSe: H-TaTe$_2$ heterostructure. InSe and H-TaTe$_2$ form an n-type SB with the height of 0.44 eV. The calculated transfer curve of H-TaTe$_2$ CS-FET is plotted in figure 4(c). It can be seen that it exhibits a lowest SS of 58 mV/decade in the V$_G$ region between -0.9 V and -0.8 V. The ON/OFF current ratio is over $10^8$ in the V$_G$ sweeping region from -1.1 V to -0.4 V at V$_{SD}$=0.74 V.

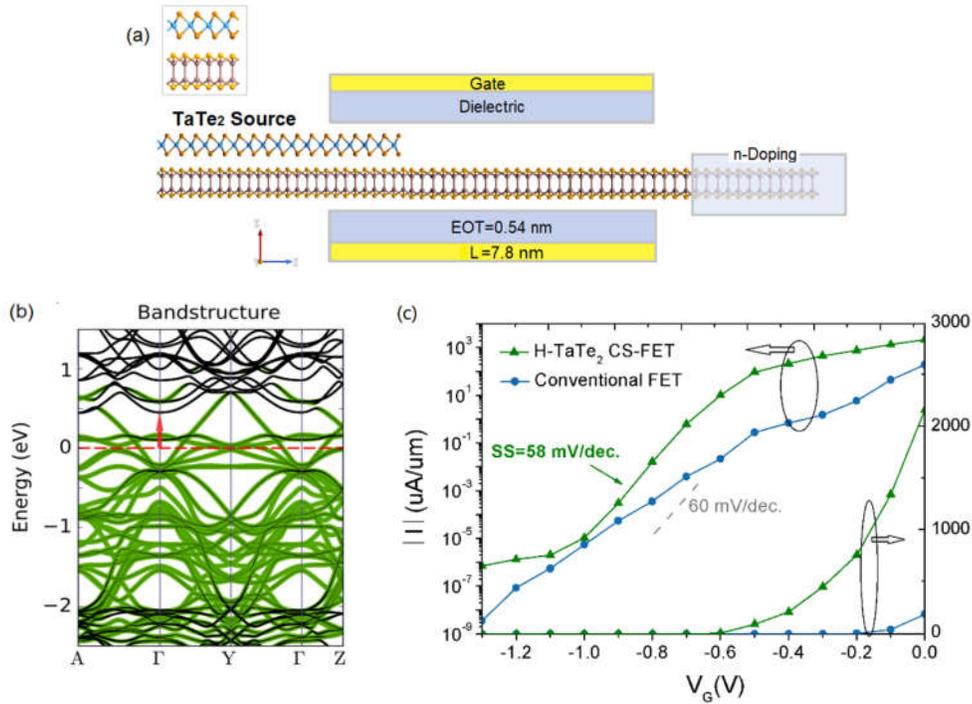

**Fig.4** (a) Side view of n-type InSe CS-FET with TaTe$_2$ source. The inset shows the relaxed atomic structure, the orange and blue balls represent Ta and Te atoms, respectively. (b) Band structure of InSe/TaTe$_2$ heterojunction. The electronic states contributed from H-TaTe$_2$ and monolayer InSe are marked by green and black lines, respectively. The Fermi level is set to zero. (c) Transfer characteristic of H-TaTe$_2$ CS-FET, I-V cure is displayed in the logarithmic and linear scales.

Figure 5(a, b) show the device DOSs in the ON and OFF states of the device, respectively. The channel barrier can be effectively modulated by gate voltage, because of the weak pinning of the FL in the vdW heterostucture. Accordingly, the transmission spectrums (figure 5(c)) of the device show significant contrast of the transmission probability at $E_F(S)$ between the ON and OFF states due to the change of the channel

barrier. Again, the gate-tunable SB can be seen from figure 5(d) which shows the atomic DOSs projected to materials in different regions of the device. The desired super-exponentially decreasing n(E) relation, which contributes to the low SS below 60 mV/decade, can be seen from the zoom-in subplots of D(E) and n(E) near the $V_G$ where the smallest SS occurs.

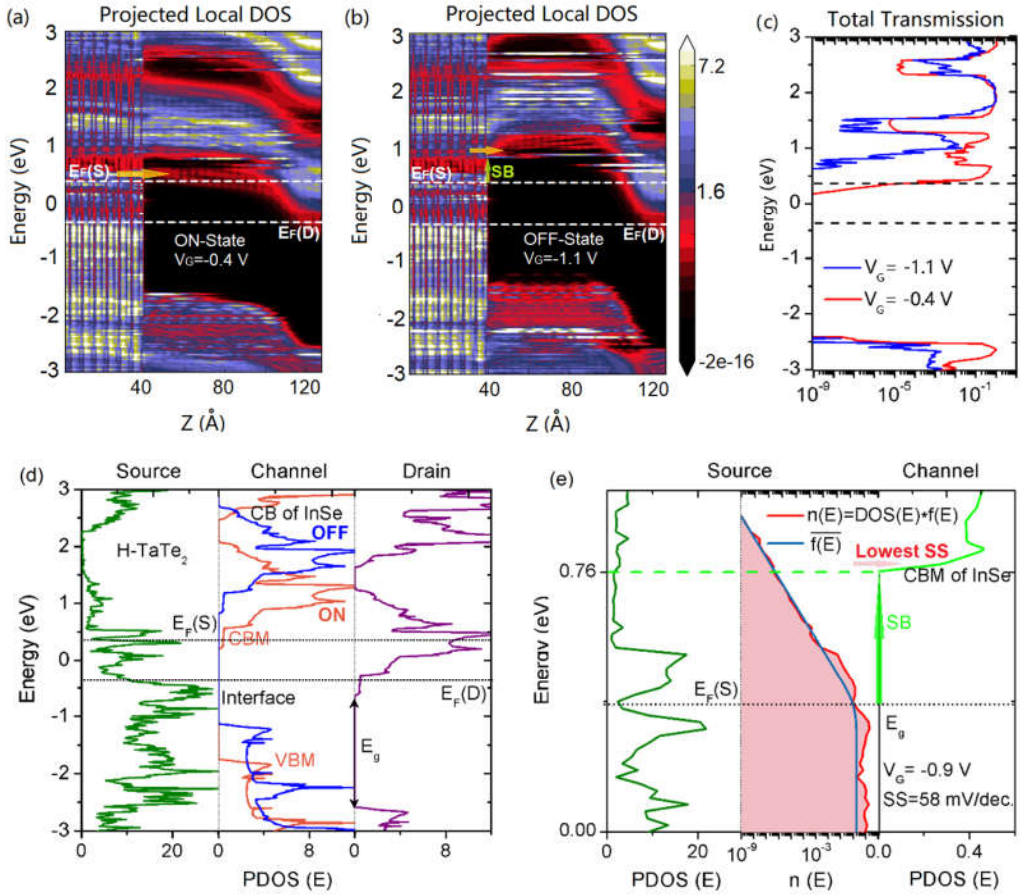

Fig. 5 PLDOS of H-TaTe$_2$ CS-FET, (a) ON- and (b) OFF-State. (c) Comparison of transmission spectrums for ON- and OFF-states. (d) Atomic DOSs projected to H-TaTe$_2$ as the source, InSe in the channel and InSe in the drain in the ON and OFF states of the device, respectively. (e) HC-DOS(E) and n(E) in the steepest-state of SS=58 mV/decade.

**Conclusion**

The application potentials of 2D vdW MS heterostructures in steep-slope CS-FETs have been studied. According to the DOS(E) relations and the nonchemical natures of the interlayer interactions of various 2D metallic materials, such as graphene, the selected TMDs and MXenes, we anticipate that a plethora of 2D vdW materials can serve as the desired CS materials. As test cases, high-mobility monolayer InSe based n-type FETs with graphene and H-TaTe$_2$ CSs have been investigated. Both FETs show low SS values below 60 mV/decade. The suppression of thermal tail contribution to the OFF current has been attributed to the gate-tunable channel barriers as well as the super-exponentially decreasing n(E). This work provides new opportunities for low-power electronics based on 2D materials.

**Methods**

Geometry optimizations are performed based on the periodic supercell method. GGA functional of the Perew-Burke-Ernzerhof (PBE) flavor is used. The atomic positions are fully relaxed until the energy difference among three consecutive steps is below 10$^{-3}$ eV and the residual forces are below 0.05 eV/Å. Van der Waals correction is done by the Grimme DFT-D2 method.

Quantum transport simulations are performed by the Atomistix ToolKit (ATK)[38] package. TB09LDA based meta-GGA (MGGA) method is used. Transport properties of different transistors are simulated by the DFT and Non-quilibrium Green's function (NEGF) methods. The device current at a given $V_G$ and $V_{SD}$ is calculated by the Landauer-Büttiker formula:[39]

$$I = \frac{2e}{h} \int_{-\infty}^{+\infty} \{T(E)[f(E-E_F(S)) - f(E-E_F(D))]\} dE \qquad (1)$$

where T(E) is the transmission probability, f(E) is the Boltzmann distribution function, $E_F(S)$ and $E_F(D)$ are the Fermi energies of source and drain. The double-zeta plus polarization (DZP) basis set is employed. Monkhorst–Pack method of 1×15×100 grid points is adopted to simulate the transport properties. The temperature is set to 300 K. The cutoff of real-space mesh is 150 Rydberg.

**Conflict of interest**

The authors declare no conflict of interest.

**Acknowledgements**

J.L., J.P. and Y.G. contributed equally to this work. This work was partly supported by National Natural Science Foundation (No. 61704096, 11464032), Beijing Natural Science Foundation (No. 4164087), SuZhou-Tsinghua innovation leading program (No. 2016SZ0102), Brain-Science Special Program of Beijing (No. Z181100001518006). Computational resources were provided by the "Explorer 100" cluster system of Tsinghua National Laboratory for Information Science and Technology as well as the Supercomputing Center of Wuhan University.